\documentclass{moriond}

\def\Journal#1#2#3#4{{#1} {\bf #2}, #3 (#4)}

\def\PRD{{\em Phys. Rev.} D}
\def\ZPC{{\em Z. Phys.} C}

\def\be{\begin{equation}}
\def\ee{\end{equation}}
\def\bea{\begin{eqnarray}}
\def\eea{\end{eqnarray}}

\newcommand{\Photo}{}

\begin{document}
\vspace*{4cm}
\title{How uncertain are model predictions for the muon content of extensive air showers}

\author{Sergey Ostapchenko}

\address{Universit\"at Hamburg, II Institut f\"ur Theoretische
Physik, 22761 Hamburg, Germany}

\maketitle\abstracts{The relation of model predictions for muon content of extensive air showers (EAS) to particular properties of hadron-air interactions
is discussed. Further, using a new Monte Carlo generator of cosmic ray interactions, QGSb, the relevance of particular interaction mechanisms to the predicted EAS muon number is studied in some detail. The corresponding constraints imposed by accelerator measurements are discussed as well.}

\section{Introduction}
Experimental studies of  very high energy cosmic rays   are performed by
indirect methods: reconstructing properties of primary
cosmic ray (CR) particles from measured characteristics of
  extensive air showers (EAS) -- nuclear-electromagnetic cascades initiated by
  interactions of primary CRs in the atmosphere. 
   Therefore, 
  such studies imply a   modeling of EAS development
  by    numerical tools, which involves a treatment  of interactions with air nuclei of both primary CRs and of secondary hadrons produced in the subsequent cascades of nuclear interactions in the atmosphere by Monte Carlo (MC) generators of hadronic interactions. \
 While such    MC generators  have been  improved substantially over the past decades, 
 they appeared  unable
 to describe consistently  a number of experimental observations.
 One particular example is the so-called  ``muon puzzle''~\cite{alb22}: the number of muons  $N_{\mu}$ measured by EAS arrays at ground level proved to exceed significantly the respective predictions of  EAS simulations.

Model predictions for  $N_{\mu}$ are 
closely correlated with the corresponding results for the moment
$\langle x_E^{\alpha_{\mu}} n_{\rm stable}^{\pi {\rm -air}}\rangle$ of the energy distribution
 of ``stable'' secondary hadrons~\cite{ost24d}  [(anti)nucleons, kaons, and charged pions] in
 pion-air interactions, defined as
 \begin{equation}
 \langle x_E^{\alpha_{\mu}} n_{\rm stable}^{\pi {\rm -air}}(E_0)\rangle 
 = \sum _{h={\rm stable}}  \int \!dx_E\; x_E^{\alpha_{\mu}}\,
 \frac{dn^{h}_{\pi {\rm -air}}(E_0,x_E)}{dx_E}\,.
 \label{eq:x*N}
\end{equation}
The summation in the right-hand-side of Eq.\ (\ref{eq:x*N}) is performed
over  secondary hadrons having significant chances to interact in the atmosphere,
  instead of decaying \footnote{Obviously, such a definition of stable hadrons is energy dependent, e.g., $K^{\pm}$ can be regarded as being stable above 1 TeV only.}, 
 $dn^{h}_{\pi {\rm -air}}/dx_E$ is their distribution, with respect to the 
energy fraction $x_E$ taken from the parent pion (upon decays of short-lived hadrons),
and $\alpha_{\mu}\simeq 0.9$   is the characteristic exponent for the
dependence of  $N_{\mu}$ on the primary energy. As is obvious from Eq.\ (\ref{eq:x*N}), 
$\langle x_E^{\alpha_{\mu}} n_{\rm stable}^{\pi {\rm -air}}\rangle$ is dominated by
forward production of stable hadrons in pion-air collisions, which is also the case
for   $N_{\mu}$ since more energetic secondaries initiate powerful enough subcascades
in the atmosphere and thereby contribute stronger to EAS muon content.
On the other hand,   $N_{\mu}$ depends rather weakly on the energy dependence of the
multiplicity of pion-air interactions, in the very high energy limit, since the
energy rise of the multiplicity is driven  by hadron production in the
 central rapidity range corresponding to small energy fractions  $x_E$,
 with the contributions of such hadrons to  Eq.\ (\ref{eq:x*N}) being
  damped \footnote{This is not the case for $\pi$-air interactions
at relatively low energies, $E_0 < 100$ GeV, where the central rapidity range
corresponds to relatively large $x_E> 0.1$.} by the 
 factor $x_E^{\alpha_{\mu}}$.
Since   $\alpha_{\mu}$ is close to unity, the moment
 $\langle x_E^{\alpha_{\mu}} n_{\rm stable}^{\pi {\rm -air}}\rangle$ 
 can be approximated by the average
fraction of the parent pion energy   taken by all stable secondary
hadrons, 
 which is approximately
equal to unity minus the energy fraction taken by neutral pions.
Therefore, to enhance the predicted  EAS muon content, one has to modify
the energy balance between stable secondary hadrons and neutral pions,
in favor of the former.

Generally there exists no viable theoretical approach which would predict forward yields of charged pions, kaons, or (anti)nucleons
 to rise with   collision energy. Therefore, any potential modifications of interaction models, aiming at predicting a higher $N_{\mu}$, are seriously constrained by accelerator data at fixed target energies. In the following, we analyze the corresponding uncertainties, using a new MC generator of CR interactions, QGSb~\cite{ops25,ops26}.

\section{Uncertainties for the predicted EAS muon content}
\begin{figure*}[t]
\centering
\includegraphics[height=2.5cm,width=0.4\textwidth]{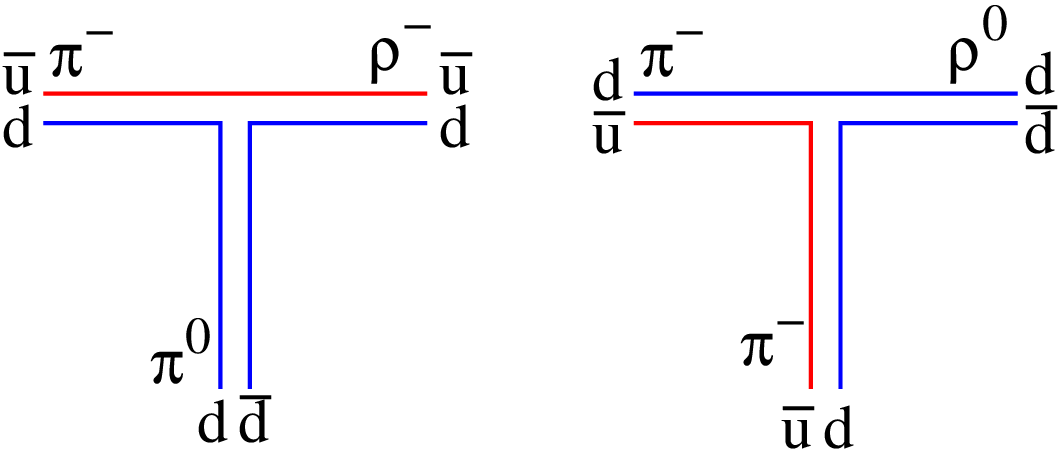}
\caption{Schematic view of the contribution  of pion exchange process to 
forward production of  $\rho$ mesons in pion-proton (pion-nucleus) interactions.
At a  microscopic level, the process is mediated by
 a creation of  light quark-antiquark pairs from the vacuum ($\bar dd$ in the figure).}
\label{fig:piex}       
\end{figure*}

A particular mechanism influencing substantially  the predicted  $N_{\mu}$ is the forward production of $\rho$ mesons in pion-air collisions, resulting from the pion exchange process~\cite{ost13}. As shown schematically in Fig.\ \ref{fig:piex}, the  %
incident pion converts into a $\rho$ meson, upon emitting a virtual pion, $\pi^{\pm}\stackrel{(\pi^{\pm})}{\rightarrow}\rho ^0$ or 
$\pi^{\pm}\stackrel{(\pi^0)}{\rightarrow}\rho ^{\pm}$, while  the virtual pion interacts with the target proton (nucleus). At a microscopic level, this corresponds to a creation of a light quark-antiquark pair ($\bar uu$ or $\bar dd$) from the vacuum; a quark (antiquark) from the pair combines with a valence  antiquark (quark) of the original pion to form the final  $\rho$ meson, while the remaining   antiquark (quark)  of the pair couples to a slowed down valence quark (antiquark), forming the virtual pion.

Neglecting this process, in nondiffractive  pion-air interactions, the energy is shared between secondary charged and neutral pions approximately \footnote{Decays of $\eta$ mesons lead to a slightly higher energy fraction given to neutral pions.} as 2:1, upon decays of short-lived hadrons, by virtue of the isospin symmetry. On the other hand, in the pion exchange process, this proportion changes to 3:1, upon decays of  $\rho$ mesons: $\rho^{\pm}\rightarrow \pi^{\pm}\pi^0$,  $\rho^{0}\rightarrow \pi^+\pi^-$. Consequently, a larger energy fraction is kept in the hadronic cascade in the atmosphere, instead of going into the electromagnetic ``sink'' via $\pi^0$ decays, thereby enhancing the calculated  $N_{\mu}$.

\begin{figure*}[t]
\centering
\includegraphics[height=4.5cm,width=0.7\textwidth]{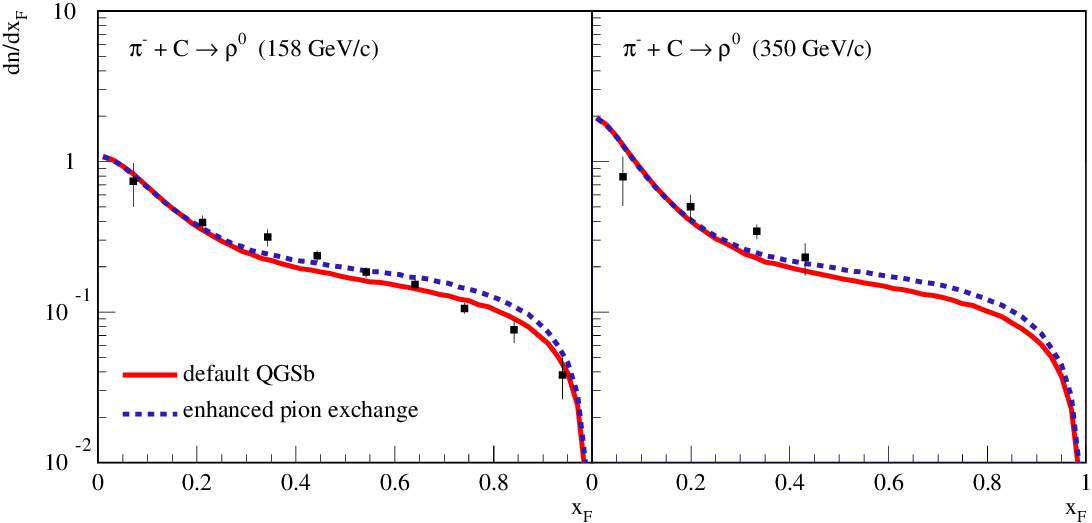}
\caption{Feynman $x$ distributions, $dn/dx_{\rm F}$, of $\rho^0$ mesons in 
center-of-mass (c.m.) frame,
 for $\pi^-C$ collisions at 158 GeV/c (left) and 350 GeV/c (right),
 as calculated using the default QGSb or by considering a higher rate of  pion exchange process --
red solid and  blue dashed lines, respectively, compared to NA61/SHINE
  data   (points).}
\label{fig:na61-rho}       
\end{figure*}

\begin{figure*}[t]
\centering
\includegraphics[height=4.5cm,width=0.7\textwidth]{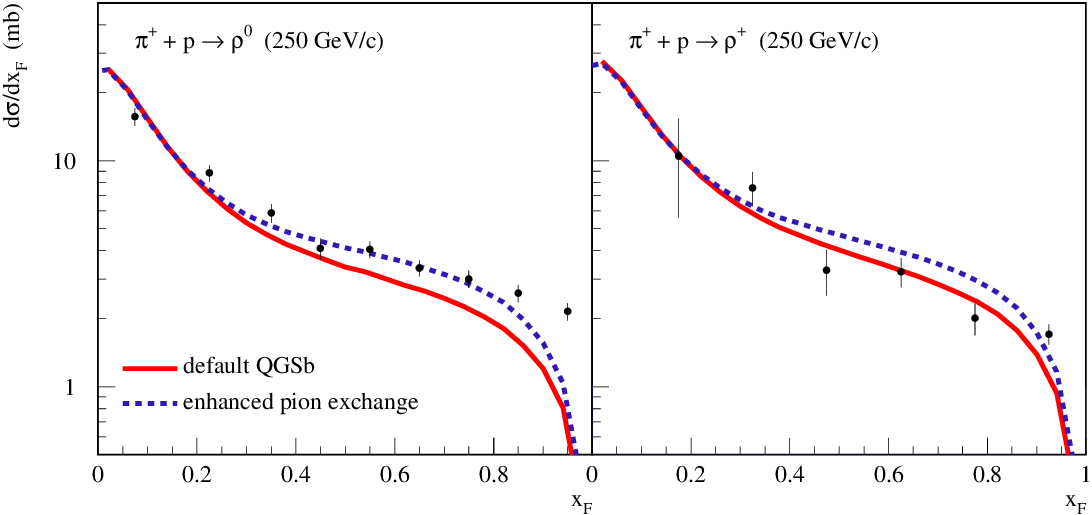}
\caption{$x_{\rm F}$-dependence of production cross sections in c.m.\ 
frame for $\rho^0$ (left) and $\rho^+$ (right), 
for $\pi^+p$ 
 collisions at 250 GeV/c, compared to EHS-NA22 data  (points).
 The notations for the  lines are the same as in  Fig.\ \ref{fig:na61-rho}.}
\label{fig:na22-rho}       
\end{figure*}

However, the rate of pion exchange is seriously constrained by accelerator measurements of  $\rho$ meson production in pion-proton and pion-nucleus collisions, notably, by the NA61/SHINE experiment~\cite{adu17}.  The results
of the QGSb model for $\rho^0$ spectra in pion-carbon collisions are shown by solid lines  in Fig.\ \ref{fig:na61-rho}, in comparison to NA61/SHINE data.
Yet measurements of  $\rho$ mesons in pion-proton interactions by the EHS-NA22 experiment~\cite{aga90}   seem to favor a somewhat higher rate of virtual pion emission, as one can see in  Fig.\ \ref{fig:na22-rho}. We can improve the agreement with the  EHS-NA22 data by %
enlarging the pion exchange contribution, the obtained  $\rho$ meson spectra being shown in  Fig.\ \ref{fig:na22-rho} by dashed lines. While such a modification creates a tension with the  data of  NA61/SHINE (cf.\ dashed lines in Fig.\ \ref{fig:na61-rho}), 
 it gives rise to 1\% only enhancement of  $N_{\mu}$.

\begin{figure*}[p]
\centering
\includegraphics[height=2.2cm,width=0.4\textwidth]{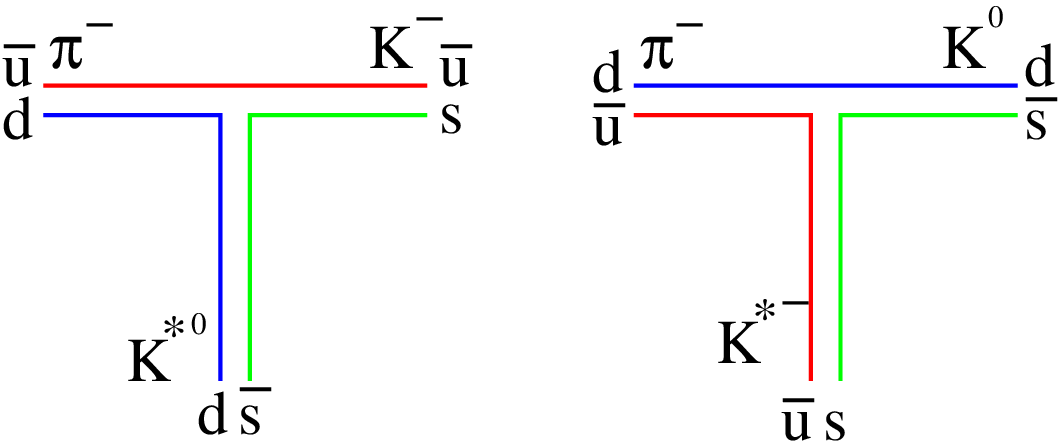}
\caption{Schematic view of the contribution  of the fragmentation of valence (anti)quarks, proceeding via  a creation of  $\bar ss$  pairs from the vacuum,  to forward production of kaons in pion-proton (pion-nucleus) interactions.}
\label{fig:K*}       
\end{figure*}

\begin{figure*}[p]
\centering
\includegraphics[height=5.9cm,width=0.7\textwidth]{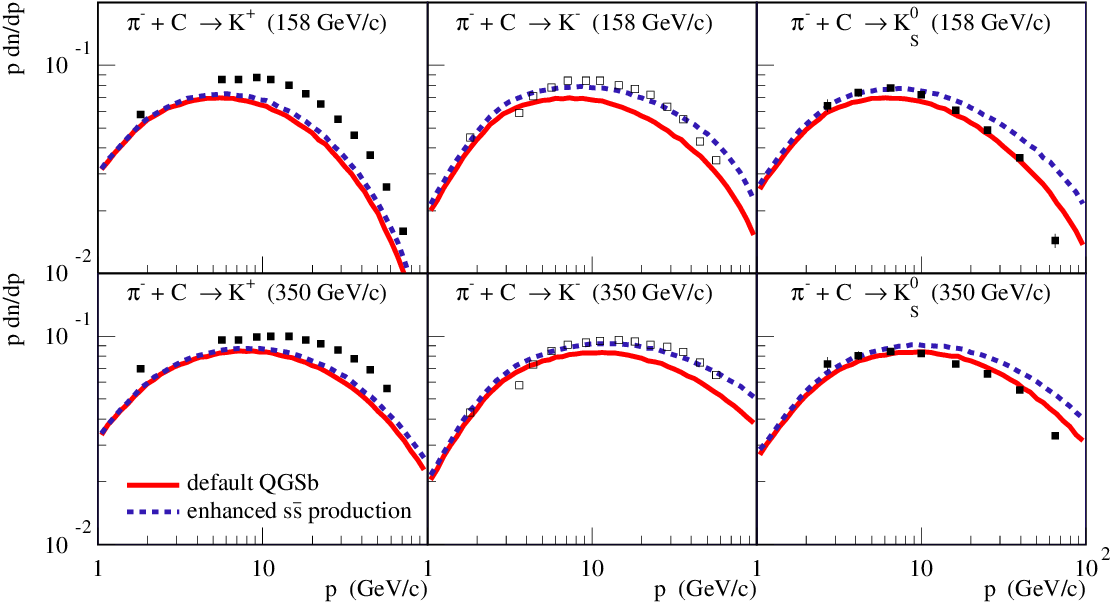}
\caption{Laboratory (lab.) momentum distributions of   $K^{+}$ (left panels),    $K^{-}$ (middle panels),
   and $K_{\rm S}^0$ (right panels) produced in $\pi^-$C collisions at 158 GeV/c (top row)
    and  350 GeV/c (bottom row), compared to NA61/SHINE data (points).
  The calculations performed using  the default QGSb or by considering a higher probability for  $\bar ss$  pair creation from the vacuum  are shown, 
  respectively, by red solid  and blue dashed lines.}
\label{fig:na61-kaon}       
\end{figure*}

Alternatively, we may try to enlarge the predicted EAS muon content by enhancing forward production of kaons and (anti)nucleons in pion-air interactions.
In the QGSb model, forward spectra of kaons are very sensitive to fragmentation of valence (anti)quarks of the pion, proceeding via a creation of $\bar ss$ quark-antiquark pairs from the vacuum, the process shown schematically in Fig.\ \ref{fig:K*}. Qualitatively, this is similar to the above-discussed pion exchange contribution: the incident pion converts into a kaon, upon emission of a virtual
$K^*$ meson,
$\pi^{\pm}\stackrel{(K^{*\pm})}{\rightarrow}K ^0_{\rm L/S}$ or 
$\pi^{\pm}\stackrel{(K^{*0})}{\rightarrow}K ^{\pm}$.
The results of the QGSb model for kaon spectra, for pion-carbon collisions, are shown by solid lines in  Fig.\ \ref{fig:na61-kaon}, in comparison to     %
  \begin{figure*}[p]
\centering
\includegraphics[height=4.5cm,width=0.7\textwidth]{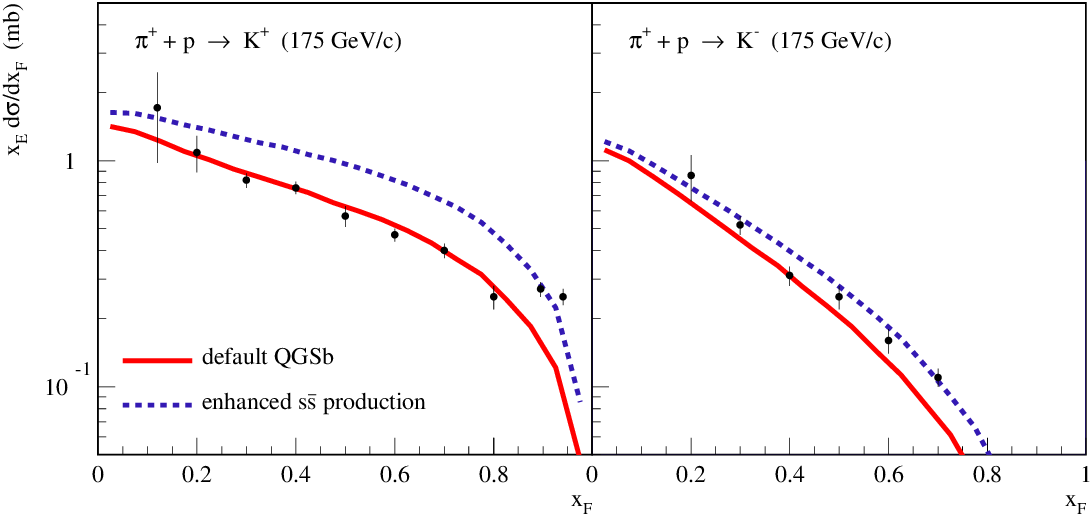}
\caption{$x_{\rm F}$-dependence of transverse momentum $p_t$ integrated
 invariant cross sections in c.m. frame, $x_E d\sigma /dx_{\rm F}$ ($x_E=2E/\sqrt{s}$), for
$K^+$ (left) and  $K^-$ (right) production in  $\pi^+p$ collisions at 175 GeV/c, 
compared to experimental data  (points).
 The notations for the  lines are the same as in  Fig.\ \ref{fig:na61-kaon}.}
\label{fig:barton-kaon}      
\end{figure*}%
\begin{figure*}[p]
\centering
\includegraphics[height=4.5cm,width=0.46\textwidth]{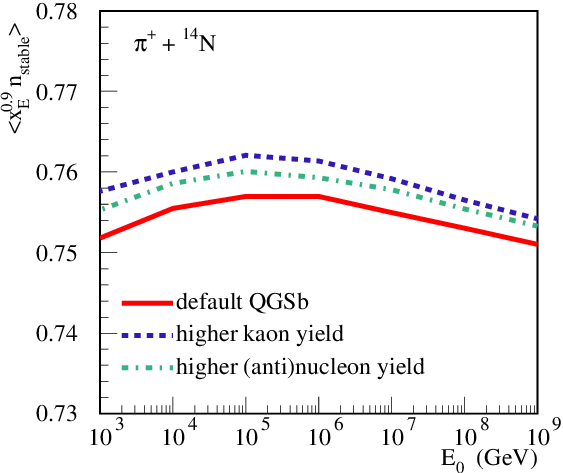}
\hfill
\includegraphics[height=4.5cm,width=0.46\textwidth]{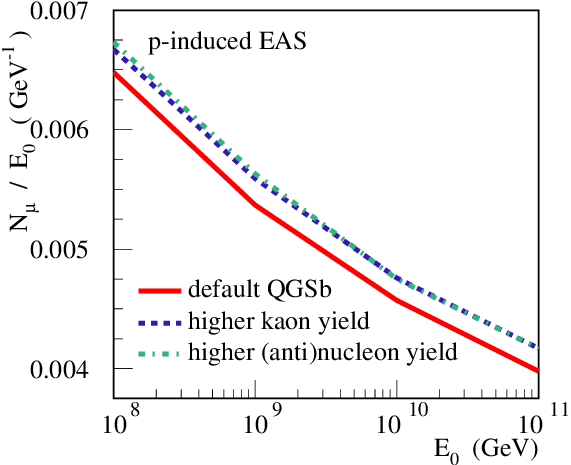}
\caption{Lab.\ energy dependence of the  moment
$\langle x_E^{0.9} n_{\rm stable}^{\pi {\rm -N}}\rangle$ (left)
and  primary energy dependence of   muon number  $N_{\mu}$ ($E_{\mu}>1$ GeV)
 at sea level, for proton-initiated vertical EAS (right),
 as calculated with  the default QGSb,
 using a higher probability for  $\bar ss$   pair creation from the vacuum, or by considering a creation of   $\bar u\bar u$-$uu$ and  $\bar d\bar d$-$dd$ diquark-antidiquark pairs 
  -- red solid, blue dashed, and green  dash-dotted  lines, respectively.}
 \label{fig:nmu-kaon}       
\end{figure*}%
NA61/SHINE data~\cite{adh23}. While the model results agree with the measurements of $K ^0_{\rm S}$, the yields of charged kaons appear to be underestimated. 
To improve the agreement for the latter, we may choose a higher probability for  $\bar ss$   pair creation from the vacuum, for the  fragmentation of valence (anti)quarks of the pion. The so-obtained kaon spectra shown by dashed lines in  Fig.\ \ref{fig:na61-kaon} go through the data points for $K^-$, while being noticeably above the measurements for  $K ^0_{\rm S}$ and still underestimating the $K^+$ yield. \footnote{It is noteworthy that  $K^+$ production in $\pi^-$-air interactions is of smaller importance for  $N_{\mu}$ predictions, compared to 
 $K^-$ and $K ^0_{\rm L/S}$, because of much softer  $K^+$  spectrum. Obviously the relative roles of  $K^+$  and  $K^-$ interchange in case of $\pi^+$-air collisions.} Yet such a modification creates a serious tension with other measurements of kaon production in pion-proton and pion-nucleus collisions. As an example, a  comparison of the
 calculated charged kaon spectra to the ones measured in pion-proton
 interactions~\cite{bre82} is shown in  Fig.\ \ref{fig:barton-kaon}.

The considered modification gives rise to 0.4--1\% enhancement of the moment
 $\langle x_E^{0.9} n_{\rm stable}^{\pi {\rm -N}}\rangle$, for pion-nitrogen interactions, as demonstrated in  Fig.\ \ref{fig:nmu-kaon}~(left). Since this     enhancement impacts the hadronic cascade development for all the cascade branchings, in the energy range where kaons can be regarded as being stable,
 one obtains an order of magnitude larger  enhancement for the calculated  number of muons ($\simeq 5$\%):
 \begin{equation}
 \frac{\Delta N_{\mu}}{N_{\mu}}\propto \left[\frac{\Delta 
 \langle x_E^{\alpha_{\mu}} n_{\rm stable}^{\pi {\rm -air}}\rangle}
 {\langle x_E^{\alpha_{\mu}} n_{\rm stable}^{\pi {\rm -air}}\rangle}\right]^{n_{\rm steps}}.
 \end{equation}
 Here the number of relevant cascade branchings $n_{\rm steps}$ refers to the cascade development at sufficiently high energies, where kaons preferentially interact, instead of decaying.
 
\begin{figure*}[t]
\centering
\includegraphics[height=2.2cm,width=0.85\textwidth]{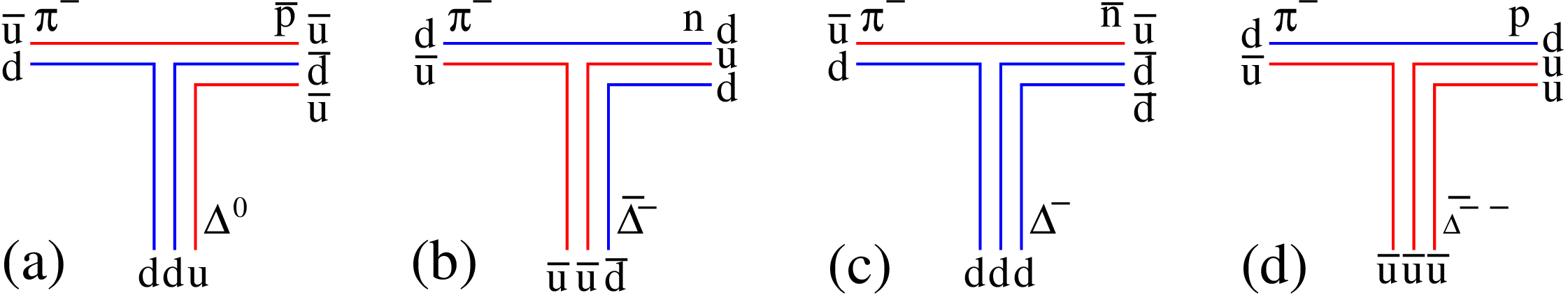}
\caption{Schematic view of the contribution  of the fragmentation of valence (anti)quarks, proceeding via  a creation of diquark-antidiquark  pairs from the vacuum,  to forward production of  (anti)nucleons in pion-proton (pion-nucleus) interactions.}
\label{fig:delta}       
\end{figure*}

\begin{figure*}[t]
\centering
\includegraphics[height=7cm,width=0.7\textwidth]{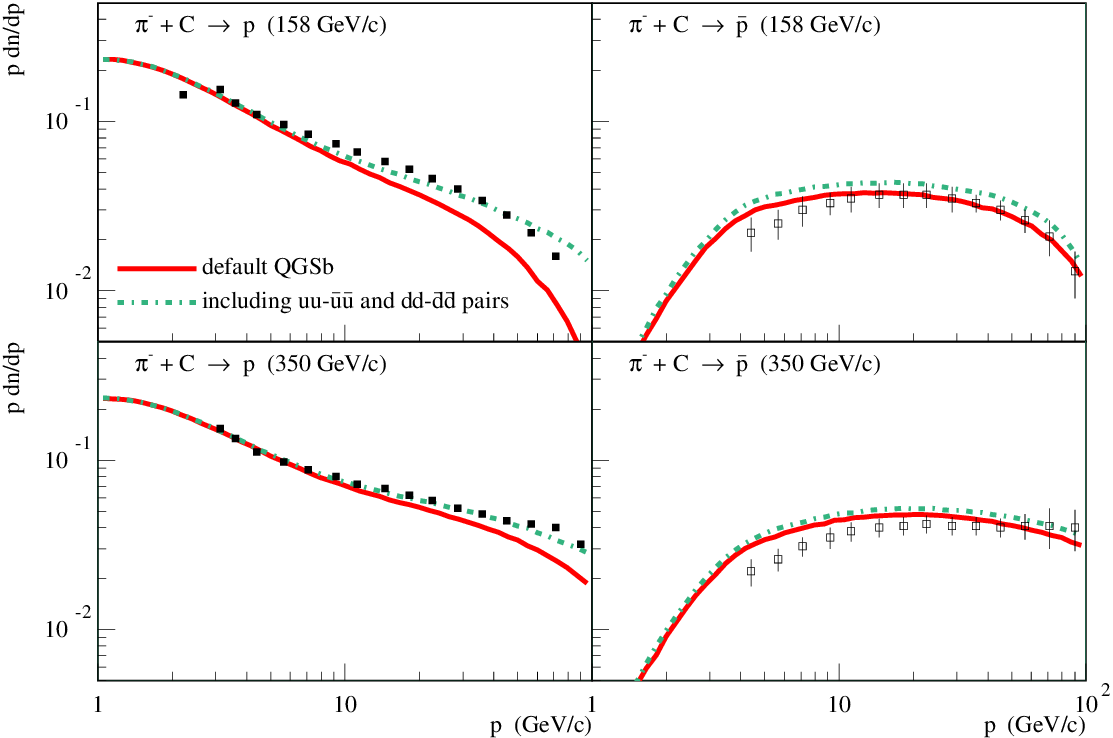}
\caption{Lab.\ momentum distributions of   protons (left panels)  and antiprotons (right panels), produced in $\pi^-$C collisions at 158 GeV/c (top row)
    and  350 GeV/c (bottom row), compared to NA61/SHINE data  (points).
  The calculations performed using the  default QGSb or by
including a creation of   $\bar u\bar u$-$uu$ and  $\bar d\bar d$-$dd$  pairs from the vacuum   are shown, 
  respectively, by red solid  and green dash-dotted  lines.}
\label{fig:na61-proton}       
\end{figure*}

\begin{figure*}[t]
\centering
\includegraphics[height=4.cm,width=0.7\textwidth]{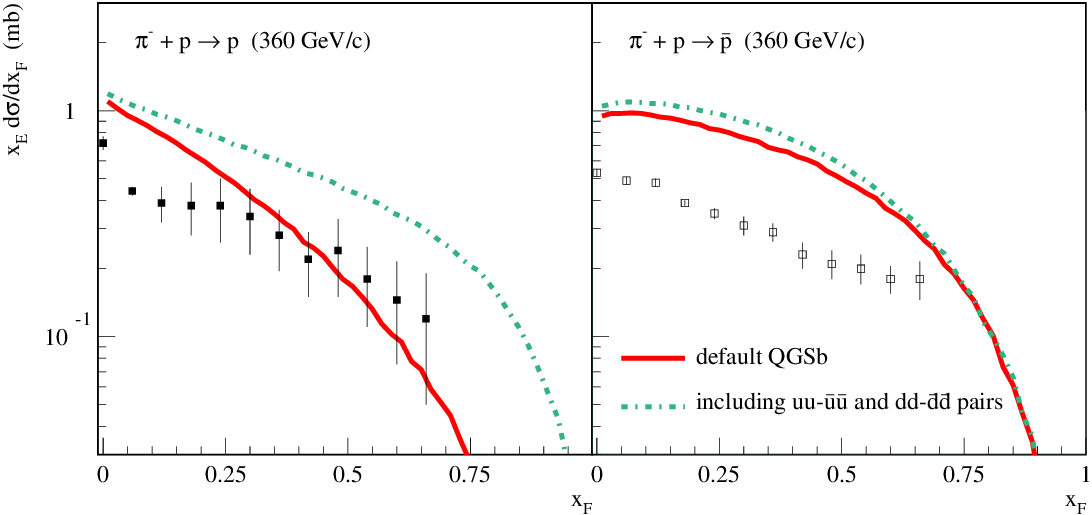}
\caption{$x_{\rm F}$-dependence of $p_t$-integrated
 invariant cross sections in c.m. frame, $x_E d\sigma /dx_{\rm F}$, for
  proton (left) and antiproton (right) production in $\pi^-p$ 
 collisions at 360 GeV/c, compared to LEBC-EHS data  (points). 
The notations for the  lines are the same as in  Fig.\ \ref{fig:na61-proton}.}
\label{fig:lebc-proton}       
\end{figure*}

 Let us finally consider a more abundant forward production of (anti)nucleons in pion-air interactions. 
In the QGSb model,   forward spectra of (anti)nucleons in pion-proton and pion-nucleus collisions are governed by fragmentation of valence (anti)quarks of the pion, proceeding via a creation of $\bar u\bar d$-$ud$ diquark-antidiquark pairs from the vacuum, the process shown schematically in Fig.\ \ref{fig:delta}~(a,b).
Qualitatively, this corresponds to an emission of virtual $\Delta$ (anti)baryons by the incident pion:
$\pi^-\stackrel{(\Delta ^0)}{\rightarrow}\bar p$ and 
$\pi^-\stackrel{(\bar \Delta ^-)}{\rightarrow}n$.
 The results of the QGSb model for the spectra of protons and antiprotons
 produced in $\pi^-$C collisions are shown by solid lines in  Fig.\ \ref{fig:na61-proton}, in comparison to   NA61/SHINE data~\cite{adh23}. 
 Here we observe a reasonable agreement with the measurements of antiproton
 production and an underestimation of forward proton yield.
 Pursuing the analogy with the  $\Delta$ (anti)baryon emission and invoking the
 isospin symmetry arguments, one may argue that emission of other kinds of
  $\Delta$ (anti)baryons should be considered as well,
  as shown in  Fig.\ \ref{fig:delta}~(c,d), i.e., one has to take
  into account the creation of  $\bar u\bar u$-$uu$ and  $\bar d\bar d$-$dd$ diquark-antidiquark pairs from the vacuum, with the same probability as for $\bar u\bar d$-$ud$   pairs. Implementing such a modification, we arrive to a reasonable agreement with the   NA61/SHINE data, both for $p$ and $\bar p$
  (cf.\ dash-dotted lines in   Fig.\ \ref{fig:na61-proton}). However, for the
  production of protons and antiprotons in pion-proton interactions, this leads to a factor   two overestimation of the corresponding spectra, compared to measurements by the LEBC-EHS experiment~\cite{agu87}, as one can see in  Fig.~\ref{fig:lebc-proton}.

This modification gives rise to $\simeq 0.5$\% enhancement of the moment
 $\langle x_E^{0.9} n_{\rm stable}^{\pi {\rm -N}}\rangle$ and up to $\simeq 6$\%
 larger  $N_{\mu}$, compared to the default QGSb results, as shown by dash-dotted lines   in  Fig.~\ref{fig:nmu-kaon}. Here, contrary to the kaon case discussed above, the considered  enhancement of forward (anti)nucleon production impacts every step of the hadronic cascade development since the number of relevant cascade branchings is not limited by decay processes.
 
To summarize our analysis, the muon content of extensive air showers is largely governed by forward production of stable hadrons in pion-air interactions, which can be quantified by the moment $\langle x_E^{\alpha_{\mu}} n_{\rm stable}^{\pi {\rm -air}}\rangle$, $\alpha_{\mu}\simeq 0.9$. Serious tensions between different sets of accelerator data, regarding forward spectra of kaons and (anti)protons produced in pion-proton and pion-nucleus collisions create some uncertainty for the calibration of relevant parameters of MC generators of CR interactions. In particular, maximizing forward yields of those hadrons, within such uncertainties, one may reach up to $\simeq 10$\% enhancement for the predicted EAS muon content.

To obtain a stronger enhancement of $N_{\mu}$, one would need to assume that
forward yields of stable hadrons rise with the collision energy, which is not supported theoretically, leaving aside exotic scenario. Importantly, since any exotic mechanism would apply equally to pion-proton (pion-nucleus) and proton-proton interactions, the corresponding predictions can be discriminated by measurements at the Large Hadron Collider. In particular, an energy rise of forward kaon yields in proton-proton collisions is already disfavored by the FASER experiment~\cite{sol26}.
 
\section*{Acknowledgments}
The author acknowledges   support from  Deutsche Forschungsgemeinschaft 
(project  550225003).

\end{document}